%% file: main.tex
\documentclass[iicol]{sn-jnl}

\usepackage{graphicx}%
\usepackage{multirow}%
\usepackage{amsmath,amssymb,amsfonts}%
\usepackage{amsthm}%
\usepackage{mathrsfs}%
\usepackage[title]{appendix}%
\usepackage{xcolor}%
\usepackage{textcomp}%
\usepackage{manyfoot}%
\usepackage{booktabs}%
\usepackage{algorithm}%
\usepackage{algorithmicx}%
\usepackage{algpseudocode}%
\usepackage{listings}%

\usepackage[latin1]{inputenc}
\usepackage{wrapfig}
\usepackage{hyperref}
\usepackage{enumerate}
\usepackage{enumitem}
\usepackage{bm}
\usepackage{multirow}
\usepackage{subcaption}
\usepackage{array}
\usepackage{tabularx}
\usepackage[normalem]{ulem}
\newcolumntype{?}{!{\vrule width 1pt}}
\makeatletter
\newcommand{\onelettername}[1]{#1\aftergroup\@gobble}
\makeatother

\usepackage{xspace}

\input{macros}
\input{def-abs}

\graphicspath{{figures/}}

\usepackage[draft,silent]{fixme}
\fxsetup{theme=color,mode=multiuser}
\definecolor{fxtarget}{rgb}{0.8000,0.0000,0.0000}
\FXRegisterAuthor{todo}{TODO}{TODO}
\FXRegisterAuthor{vp}{VP}{{\tiny \color{violet}VP}}
\FXRegisterAuthor{ra}{RA}{\tiny \color{blue}Rizwan}

\theoremstyle{thmstyleone}%
\theoremstyle{thmstyletwo}%

\theoremstyle{thmstylethree}%

\begin{document}

\title[$\easyrpl$]{$\easyrpl$}
\subtitle{A web-based tool for modelling and analysis of cross-organisational workflows}

\author*[1]{\fnm{Muhammad Rizwan} \sur{Ali}}\email{mral@hvl.no}

\author[1]{\fnm{Violet Ka I} \sur{Pun}}\email{violet.ka.i.pun@hvl.no}

\author[2]{\fnm{Guillermo} \sur{Rom\'an-D\'iez}}\email{guillermo.roman@upm.es}

\affil*[1]{ \orgname{Western Norway University of Applied Sciences}, \orgaddress{ \city{Bergen}, \country{Norway}}}

\affil[2]{\orgname{Universidad Polit\'ecnica de Madrid},  \country{Spain}}


\abstract{%
Cross-organisational workflows involve multiple concurrent, collaborative workflows across different departments or organisations, necessitating effective coordination due to their interdependent nature and shared resource requirements. The complexity of designing and managing these workflows stems from the need for comprehensive domain knowledge and a unified understanding of task dependencies and resource allocation. Existing tools often fall short in facilitating effective cross-organisational collaboration and resource sharing. This paper introduces $\easyrpl$, a user-friendly web-based tool suite designed to manage cross-organisational workflows. $\easyrpl$ consists of a simulator for visualising the impact of workflow changes, a peak resource analysis tool for identifying potential resource bottlenecks, and a time analysis tool for estimating execution time. These tools assist planners with detailed insights to optimise workflow efficiency and minimise disruptions, enhancing the management of complex, interdependent workflows.}

\keywords{Cross-organisational workflows, resource sharing, workflow management, simulation, peak resource analysis, time analysis.}



\maketitle

\section{Introduction}\label{sec:intro}
A cross-organisational workflow involves concurrent, collaborative
workflows operating across different departments or multiple
organisations. These workflows are interdependent, with tasks in one
workflow often relying on the completion of tasks in other
collaborative workflows. Additionally, they may share resources for
task execution, necessitating effective coordination and
synchronisation. For example, in a supply chain management workflow,
different entities such as suppliers, manufacturers, distributors, and
retailers each have their own distinct workflows.
Efficient coordination of these entities based on task dependencies
and shared resources is required to ensure the timely and efficient delivery of products.

Designing cross-organisational workflows is inherently complex. It requires a
comprehensive understanding of multiple specific domains, as well as an
overarching view of the dependencies on shared resources and task completion
across the various collaborative workflows. This complexity makes the
optimisation and management of these workflows challenging and  error-prone.
Minor local changes can be propagated unpredictably through the system, potentially
disrupting the entire workflow network. This can lead to substantial financial
loss or, in critical supply chain contexts, significant delays and customer
dissatisfaction. In particular, in sectors such as healthcare, minor mistakes can
result in fatalities, highlighting the critical need for meticulous workflow
management.

Tools such as Process-Aware Information Systems (PAIS)~\cite{pais} and
Enterprise Resource Planning (ERP)~\cite{erp} systems have been
developed for workflow management. However, existing tools and
techniques may lack the ability to share knowledge and resources
across organisations and account for inter-organisational task
dependencies. These factors are essential for accurately managing
changes in workflows. Without the ability to effectively share
information and resources across organisational boundaries, planners
may lack a comprehensive understanding of the entire workflow
ecosystem, leading to inaccuracies in cost projection and potential
inefficiencies in resource allocation.

The lack of comprehensive domain knowledge among planners and the
absence of a unified understanding of all collaborative workflows,
especially in safety critical areas such as healthcare, emphasise the
importance of a formalism that can allow planners to model and
simulate cross-organisational workflows, which allows planners to
estimate the effect of local changes on all the collaborative
workflows prior to the actual implementation of the changes.

Our earlier work introduced a resource-sensitive formal modelling
language, $\rpl$~\cite{sbmf}, designed to model cross-organisational
workflows, sharing resources. This language employs an actor-based
concurrency model~\cite{Agha86-book}, incorporating notions of time
advancement, task completion deadlines, and explicit task
dependencies. Furthermore, a static cost analysis has been proposed to
approximate the worst execution time of workflows modelled in
$\rpl$~\cite{ALI2023102896}. The analysis is implemented as part of
the proof of concept tool $\rpltools$~\cite{10.1007/978-3-031-35257-7_18}, which also includes functionalities for modelling and simulation.

This paper presents $\easyrpl$, a tool suite with a user-friendly, simple web-based
interface, designed to assist in cross-organisational workflows management.
$\easyrpl$ comprises three tools: simulator, peak resource analysis, and time
analysis.
%
The simulator allows planners to see the potential impacts
of changes in resource availability, efficiency, and the number of concurrent
cases on metrics, such as deadline violations, execution time, and financial
costs within collaborative workflows through simulation.
Compared to the simulation module of
$\rpltools$,
$\easyrpl$ uses graphical representation, in the form of
tables and bar charts, to present detailed simulation information,
making it easier for users to visualise and understand
the results.
Peak resource analysis tool returns the maximum amount of
resources that could be allocated at the same time in a cross-organisational
workflow defined in $\rpl$, utilising the \textit{peak} resource analysis
presented in~\cite{AlbertCR18}. This helps in identifying potential resource bottlenecks
and optimising resource allocation.
Finally, the time analysis in $\easyrpl$, compared to
\cite{ALI2023102896}, allows consuming time specified as a
parametrised expression.
Furthermore, the time analysis returns an upper-bound in closed-form
by solving cost equations with the off-the-shelf solver
PUBS~\cite{AlbertAGP11a}.

The rest of the paper is organised as follows:
Section~\ref{sec:background} presents the architecture of $\easyrpl$
and describes the input workflow models in $\rpl$ with a simple example.
Section~\ref{sec:easyrpl} presents the web-interface of $\easyrpl$.
Sections~\ref{sec:simulation}, \ref{sec:peak}, and \ref{sec:time}
present the three tools, respectively.
Section~\ref{sec:relatedWork} discusses the related work.
Finally, we summarise the paper and discuss possible future work in Section~\ref{sec:conclusion}.

\section{Background}
\label{sec:background}
\input{language}

\section{$\easyrpl$}
\label{sec:easyrpl}
\input{web}

\section{Simulator}
\label{sec:simulation}
\input{simulation}

\section{Peak Resource Analysis}
\label{sec:peak}
\input{peak}

\section{Time Analysis}
\label{sec:time}
\input{time}

\section{Related Work}
\label{sec:relatedWork}
\input{related}

\section{Conclusions}
\label{sec:conclusion}

This paper introduced $\easyrpl$, a tool suite with a user-friendly web-based interface that assists planners in simulating workflow changes, analysing resource usage, and estimating execution times with accuracy and ease. The simulator allows users to visualise the impact of changes in resource availability, efficiency and concurrency level, presenting detailed information through tables and bar charts. This graphical representation aids in understanding the potential effects on key metrics such as deadline violations, execution time, and financial costs. The peak resource analysis tool helps identify potential resource bottlenecks, ensuring optimal resource allocation and preventing inefficiencies. The time analysis tool provides closed form upper-bound time estimates by solving cost equations, contributing to more precise planning and execution of workflows.

By leveraging these tools, planners can achieve a more comprehensive understanding of cross-organisational workflows, leading to improved efficiency, reduced risks of disruptions, and enhanced coordination among collaborative entities. 

\smallskip
\noindent
\textbf{{Future Work.}}
A natural extension would be an additional interface allowing users to graphically design workflows that can be translated to $\rpl$ models.
We also plan to validate the tool suite through real-world case studies in different industries to investigate its applicability in practice.
In addition, we intend to assess the performance of the three tools, in particular their scalability, efficiency and accuracy, by comparing with other workflow modelling tools.

\subsubsection*{\small Acknowledgements.} 
This work is part of the \textsc{CroFlow}
project: Enabling Highly Automated Cross-Organisational Workflow
Planning, funded by the Research Council of Norway (grant no. 326249),
and is partially funded by the Spanish MCIU, AEI and FEDER (EU)
projects PID2021-122830OA-C44 and PID2024-157044OB-C33 and by the
Comunidad de Madrid grant {TEC-2024/COM-235} (DESAF\'IO-CM). 
\bibliographystyle{sn-aps}
\bibliography{main}

\newpage
\appendix
\onecolumn
\section{BPMN representation of the running example}
\label{sec:bpmn}
\begin{figure*}[h!]
\centering
\includegraphics[width=1\textwidth]{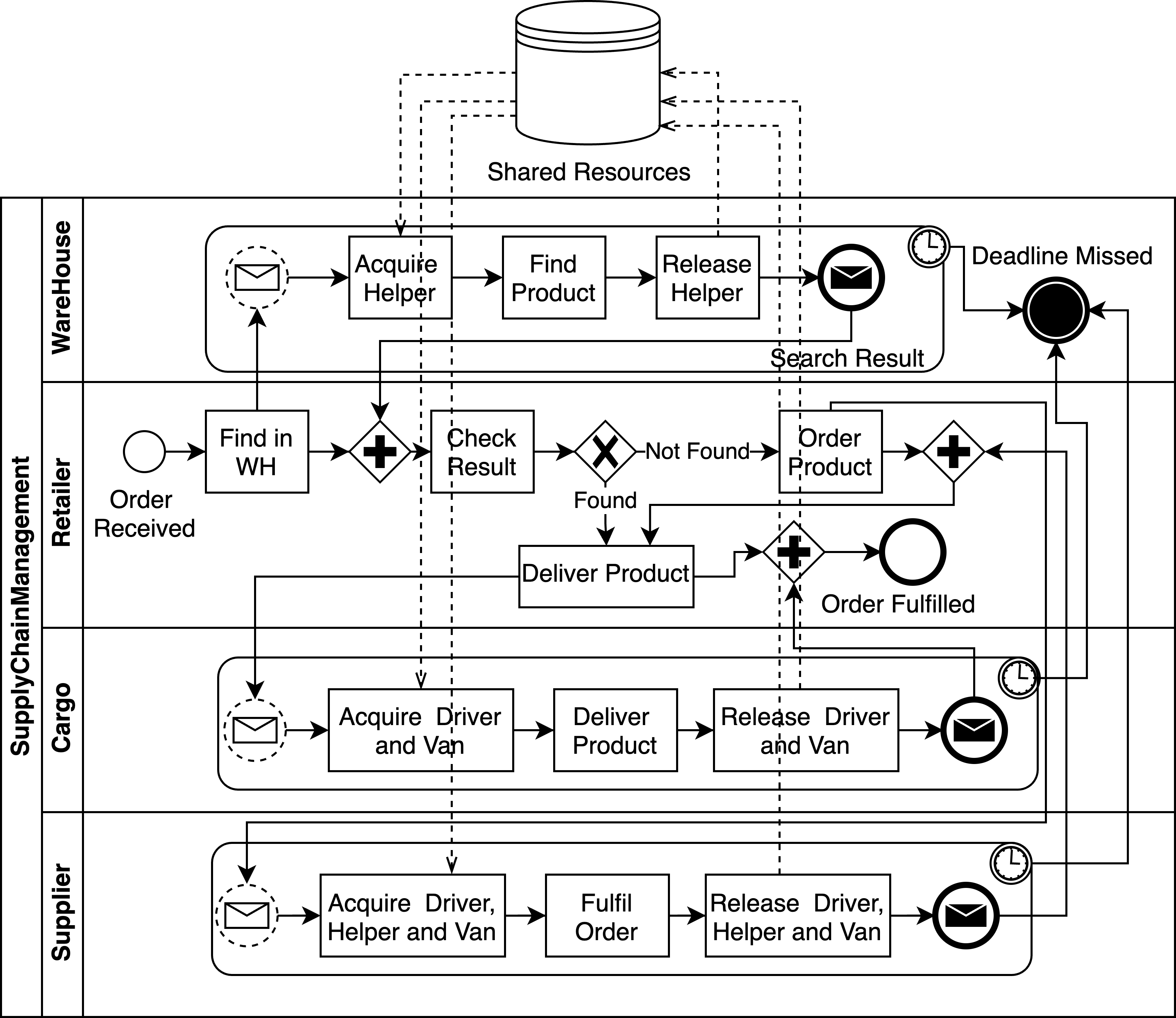}
\caption{A simple example of collaborating workflows in BPMN}
\label{fig:example}
\end{figure*}

\section{Workflow modelled in $\rpl$ of the running example}
\label{sec:rpl-code}
\input{example-code}	

\end{document}

%% file: macros.tex
\definecolor{lightgray}{gray}{0.9}
\definecolor{airforceblue}{rgb}{0.36, 0.54, 0.66}
\definecolor{bittersweet}{rgb}{1.0, 0.44, 0.37}
\definecolor{brilliantlavender}{rgb}{0.96, 0.73, 1.0}

\newcommand{\res}{\ensuremath \mathit{res}}

\def\codesize{\fontsize{9}{10}}

\newcommand{\rpl}{\mathcal{R}{\normalfont\textsc{pl}}}
\newcommand{\easyrpl}{{\normalfont\textsc{Easy}}\mathcal{R}{\normalfont\textsc{pl}}}
\newcommand{\rpltools}{\ensuremath\rpl {\rm Tool}}

\newcommand{\main}{\ensuremath\mathit{main}}

\newcommand{\UB}{\mathit{UB}}
\newcommand{\qua}{\mathit{qua}}

\newcolumntype{L}[1]{>{\raggedright\let\newline\\\arraybackslash\hspace{0pt}}m{#1}}
\newcolumntype{C}[1]{>{\centering\let\newline\\\arraybackslash\hspace{0pt}}m{#1}}
\newcolumntype{R}[1]{>{\raggedleft\let\newline\\\arraybackslash\hspace{0pt}}m{#1}}

\newcommand{\alt}{\texttt{alt}\xspace}

%% file: def-abs.tex
\lstdefinelanguage{ABS}{keywords=
{assert, switch, when, newclass,newinterface,update,foreach,this,thisDC,dyndelta,new,data,type,def,case,case2,of,cog,class,interface,extends,implements,if,else,await,get,local,Fut,return,skip,while,module,duration,duration2,now,deadline,import, export, uses, from,
suspend,delta,adds,modifies,removes,original,then,productline,features
,corefeatures,optionalfeatures,after,when,product,hasAttribute,hasMethod,root,extension,group,allof,oneof,require,exclude,original,
ifin,ifout,opt,spawn,newgroup,acquire,in,except,joins,leaves,as,subtypeOf,after,hold,
release, while, skip, interface, class, resource, cost, dl, Res,max,speed, RP, Set, Map, Rp, Rid, Unit, Bool, not, xor, and, True, Flase, Int}, 
sensitive=true, comment=[l]{//}, morecomment=[s]{/*}{*/}, morestring=[b]"}

\definecolor{codegreen}{rgb}{0,0.6,0}
\definecolor{codegray}{rgb}{0.5,0.5,0.5}
\definecolor{codepurple}{rgb}{0.58,0,0.82}
\definecolor{backcolour}{rgb}{0.96,0.96,0.96}
\definecolor{commentcolour}{rgb}{0.49, 0.45, 0.65}

\definecolor{codeblue}{rgb}{0.42,0.56,0.75}
\definecolor{framecolor}{rgb}{0,0,0}

\newcommand{\Abs}[1]{\lstinline[language=ABS,columns=fullflexible,mathescape=true,inputencoding=latin1,extendedchars,keywordstyle=\bf\codesize\sffamily,basicstyle=\codesize\sffamily]!#1!}

\newenvironment{absinline}[1][]{%
	\lstset{language=ABS,
		showstringspaces=false,
		columns=fullflexible,mathescape=true,%
		keywordstyle=\bf\sffamily,commentstyle=\sl\sffamily,%
		basicstyle=\normalsize\sffamily,inputencoding=latin1, 
		extendedchars,xleftmargin=2em,#1}}
{}

\lstdefinestyle{absstyle}{
	language=ABS,columns=fullflexible,
	mathescape=true,%
	showstringspaces=false,%
	keywordstyle=\bf\sffamily,
	numberstyle=\tiny,
	numbers=none,
	commentstyle=\tiny\sffamily,%
	basicstyle=\tiny\sffamily,
	inputencoding=latin1, 
	extendedchars,xleftmargin=2pt
}

\lstnewenvironment{absexample}{
	\noindent 
	\lstset{style=absstyle,
		basicstyle=\myttsize\sffamily,
		keywordstyle=\bfseries\color{codepurple},
		framerule=1pt,
		backgroundcolor=\color{backcolour},
		rulecolor=\color{codepurple!80!black},
		frame=tblr,
		xleftmargin=4pt,
		xrightmargin=6pt,
}}
{}

\lstnewenvironment{absexamplen}[1][]{
	\lstset{style=absstyle,
		basicstyle=\myttsize\sffamily,
		keywordstyle=\bfseries\color{codepurple},
		framerule=1pt,tabsize=2,
		backgroundcolor=\color{backcolour},
		rulecolor=\color{Green!80!black},
		frame=tblr,
		captionpos=b,
		xleftmargin=4pt,#1
}}
{}

\makeatletter
\lstnewenvironment{absexamplesm}[1][]{
	\noindent 
	\lstset{style=absstyle,
		basicstyle=\scriptsize\sffamily,
		keywordstyle=\bfseries\color{codeblue},
		framerule=0.8pt,
		backgroundcolor=\color{backcolour},
		commentstyle=\bfseries\color{commentcolour},
		rulecolor=\color{framecolor},
		frame=tblr, 
		xleftmargin=10pt,
		xrightmargin=10pt,
		numbers=left, 
		numbersep=6pt,
		#1,
	}
	\csname\@lst @SetFirstNumber\endcsname
}{%
	\csname \@lst @SaveFirstNumber\endcsname
}
\makeatother

%% file: language.tex
\begin{figure}[tp!]
\centering
\includegraphics[width=0.45\textwidth]{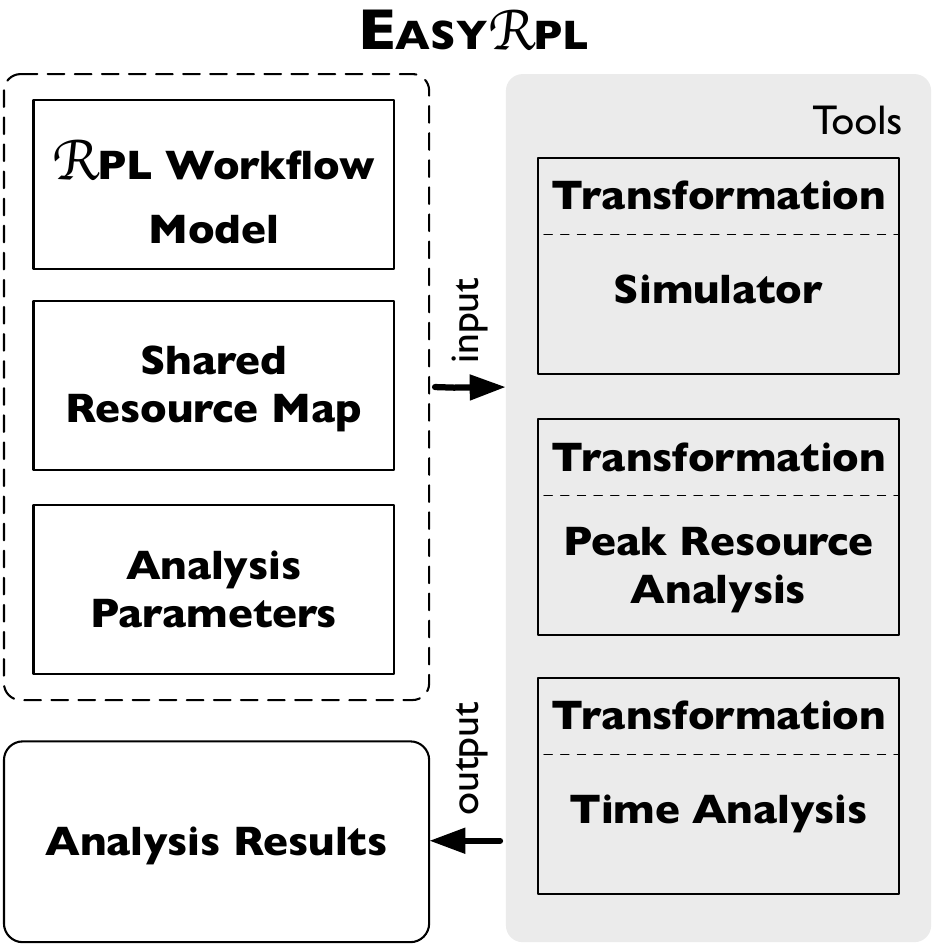}
\caption{Architecture of $\easyrpl$.}
\label{fig:architecture}
\end{figure}

Figure~\ref{fig:architecture} illustrates the architecture of
$\easyrpl$, a tool suite designed for workflow modelling and
analysis. $\easyrpl$ takes as inputs a workflow modelled in $\rpl$, a
map of shared resources along with tool-specific parameter
configurations through its web interface. Then, based on the selected
analysis, user can run one of its tools: Simulator, Peak Resource
Analysis, or Time Analysis. Each tool translates an input $\rpl$ program into
a specific ABS~\cite{johnsen2010abs} format and serves distinct
functions within the workflow analysis process.
Subsequently, $\easyrpl$ generates corresponding outputs through the
web interface, providing the simulation or analysis
results which reflect the performance and behaviour of the workflow.

In this section we present the formal modelling language $\rpl$~\cite{ALI2023102896} and explain how resources can be specified in $\rpl$. Furthermore, we use a simple example to show how the different workflows can be coupled via shared resources in $\rpl$. The detailed overview of $\easyrpl$ web interface and parameter configurations can be found in Section~\ref{sec:easyrpl}.

\subsection{Resource Sensitive Workflow Modelling Language $\rpl$}
\label{sec:rpl}
One of the inputs of $\easyrpl$ is a cross-organisational workflow
modelled in~$\rpl$, which is a formal modelling language, inspired by
ABS~\cite{johnsen2010abs}, designed for cross-organisational
workflows. $\rpl$ couples multiple workflows through shared resources
and task dependencies and supports explicit notions of task completion
deadlines and time advancement.

An $\rpl$ program consists of several components:
first, the declaration of a module name,
followed by interface and class declarations, and finally,
a main block.
Tasks in workflows are represented as methods in a class in $\rpl$ and
are executed through method invocations, which can be either
synchronous or asynchronous.
Synchronous method invocations block the caller objects until the
called methods return, whereas asynchronous calls are non-blocking,
enabling both callers and callees to run in parallel.
Each asynchronous method invocation is associated with a future, wherein
the invoked method stores the return value upon completion.
Every task may have a completion deadline, denoted by the notion
\Abs{dl},
and the commencement of the task execution may depend on the
completion of other tasks, indicated by the \Abs{after} keyword
followed by the futures associated with the depending tasks (method
calls).
Time advancement in $\rpl$ models is captured by the
\Abs{cost} statement.

Resources can be acquired and returned with statements \Abs{hold} and \Abs{release}. Resources shared across collaborative workflows are represented using a map that links each resource $r$ to a pair ($b, Q$), where $b$ is a Boolean indicating whether the resource is available, and $Q$ is a set describing the resource's qualities. These qualities include the resource's category, efficiency (in terms of experience for a human and power for a machine), and financial usage cost.

\begin{figure}[h!]
{
\begin{tabular}{rl}
[&$\cdots$~,\\
&$r_1 \mapsto$ (true, \{Van,1500,500,9\}), \\
&$r_2 \mapsto$ (false, \{Driver,7,100,1\}), \\
&$r_3 \mapsto$ (false, \{Van,2000,750,12\}), \\
&$r_4 \mapsto$ (true, \{Helper,9,50,5\}), \\
&$\cdots$~~]
\end{tabular}
}
\caption{An example of resource map in $\rpl$ shared between different departments.}
\label{fig:simple-resources}
\end{figure}

\setlength{\columnsep}{45pt}%
\begin{figure*}[tp!]
\begin{absexamplesm}
class Retailer implements Retailer {
$\label{ex:process-simple-var}~~~~$Int found = 0; 
$\label{ex:process-order}~~~~$Int process_order(Warehouse wr,Cargo cr,Supplier sup) {
$\label{ex:process-simple-var1}~~~~~~~~$Int sent = 0;
$\label{ex:process-futVar}~~~~~~~~$Fut<Int> f1; Fut<Int> f2; Fut<Int> f3;
$\label{ex:check-goods}~~~~~~~~$f1 = !check_goods(wr) after dl 10;
$\label{ex:await-check-goods}~~~~~~~~$await f1?;
$\label{ex:get-check-goods}~~~~~~~~$found = f1.get;
$\label{ex:if}~~~~~~~~$if(found == 1){
$\label{ex:if-send}~~~~~~~~~~~~$f2 = !deliver_goods(cr) after dl 170;
$\label{ex:if-await-send}~~~~~~~~~~~~$await f2?;
$\label{ex:if-get-send}~~~~~~~~~~~~$sent = f2.get;
$\label{ex:else}~~~~~~~~$} else {
$\label{ex:else-order}~~~~~~~~~~~~$f2 = !order_goods(sup) after dl 220;
$\label{ex:else-send}~~~~~~~~~~~~$f3 = !deliver_goods(cr) after f2 dl 170;
$\label{ex:else-await-send}~~~~~~~~~~~~$await f3?;
$\label{ex:else-get-send}~~~~~~~~~~~~$sent = f3.get;
$~~~~~~~~$}
$\label{ex:return-process-order}~~~~~~~~$return sent;
$~~~~$}
}
\end{absexamplesm}
\caption{A simple example of collaborating workflows in $\rpl$ -- part I.}
\label{fig:simple-rpl-1}
\end{figure*}

For example, Figure~\ref{fig:simple-resources} shows the map of the resources shared between the collaborating departments to deliver goods. For instance, $r_1$ is a van of 1500cc power with a capacity of 9 passengers and a usage cost of 500 per time and is currently free.  While $r_2$ is a driver with seven years of experience and a usage cost of 100 per time unit, and is currently busy.  Note that the quality set~$Q$ each of the resource type can be extended with additional features.
The rest of the language rather standard, and 
we refer the readers to \cite{ALI2023102896} for the full syntax and semantics.

\setlength{\columnsep}{45pt}%
\begin{figure*}[h!]
\begin{absexamplesm}
class Cargo implements Cargo {
$~~~~\label{ex:send-effort}$Int delivery_effort = 150;
$~~~~\label{ex:send-time}$Int delivery_time= truncate(delivery_effort*(100/$\$$EFFICIENCY));
$~~~~$Int deliver_goods() {
$\label{ex:send.hold}~~~~~~~~$Pair<List<Int>,Int> p = hold(list[set[ResEfficiency(70),Driver],
$~~~~~~~~~~~~~~~~~~~~~~~~~~~~~~~~~~~~~~~~~~~~~~~~$set[ResEfficiency(1500),Van]]);
$\label{ex:send.cost}~~~~~~~~$cost(delivery_time); 
$\label{ex:send.p}~~~~~~~~$p = Pair(fst(p),snd(p)*delivery_time);
$\label{ex:send.release}~~~~~~~~$release(p);
$\label{ex:send.return}~~~~~~~~$return 1;
$~~~~$}
}
\end{absexamplesm}
\caption{A simple example of collaborating workflows in $\rpl$ -- part II.}
\label{fig:simple-rpl-2}
\end{figure*}

\subsection{A Simple Example}
Figure~\ref{fig:simple-rpl-1} models in~$\rpl$ a simple order processing workflow in the $\mathtt{Retailer}$, who needs to interact with the $\mathtt{Warehouse}$, the $\mathtt{Supplier}$ and the $\mathtt{Cargo}$ team.
To process the order, the $\mathtt{Retailer}$ sends a request to the $\mathtt{Warehouse}$ to check if the goods are available, captured by the asynchronous method invocation in Line~\ref{ex:check-goods}, where the callee is indicated by first parameter of the method calls.
While waiting for the response from the $\mathtt{Warehouse}$ in Line~\ref{ex:await-check-goods}, the $\mathtt{Retailer}$ can process one of the ready tasks in the waiting queue.
After the \texttt{Warehouse} has checked the availability of the goods, the $\mathtt{Retailer}$ gets the result on Line~\ref{ex:get-check-goods}.
If the goods are available, the $\mathtt{Retailer}$ instructs the $\mathtt{Cargo}$ team to send the goods to customer through an asynchronous method invocation (Line~\ref{ex:if-send});
otherwise, the $\mathtt{Retailer}$ will asynchronously place an order for goods with the \texttt{Supplier} (Line~\ref{ex:else-order}).
As soon as the goods are received from the $\mathtt{Supplier}$, the $\mathtt{Retailer}$ sends an asynchronous request to the $\mathtt{Cargo}$ team to deliver the goods to customer in Line \ref{ex:else-send}, indicated by the future~\Abs{f2} following the \Abs{after} keyword, where \Abs{f2} is the future associated to the method call to $\mathtt{Supplier}$ in Line~\ref{ex:else-order}.
Note that method invocations in Lines~\ref{ex:check-goods}, \ref{ex:if-send} and~\ref{ex:else-order} do not have any task dependency, indicated by having no future following the~\Abs{after} keyword, but completion deadlines of ten, thirty and fifty time units, respectively.

Figure~\ref{fig:simple-rpl-2} models the cargo workflow in~$\rpl$, where a driver and a van, with efficiency in terms of at least seven years of experience and a power of 1500, respectively, are initially acquired from the resource pool in Line~\ref{ex:send.hold} and released in Line~\ref{ex:send.release} after the goods have been delivered. 
This process is modelled as time advancement, captured by the \Abs{cost} statement where \Abs{send_time} depends on the percentage of efficiency of the resources, as indicated by \$\Abs{EFFICIENCY} (see details later in Section~\ref{sec:parameter}) in Line~\ref{ex:send-time}. The \Abs{hold} statement returns a list of identifiers of the acquired resources and the sum of the usage cost per unit time, while the \Abs{release} statement takes a pair as input, comprising a list of identifiers of the acquired resources and the total financial costs of using the acquired resources over the specified time duration in the \Abs{cost} statement.

Following the main block, users need to specify the resources, preceded
by the keyword \Abs{Resources} followed by a colon (:). Resources of the same
type should be listed together, with each resource set containing qualities such
as category, efficiency and cost separated by coma (,) listed per line.
Different types of resources are differentiated by the \Abs{$\$$} symbol.
	

A graphical representation of the model in BPMN and the complete model in $\rpl$ can be found in Figure~\ref{fig:example} and Figure~\ref{fig:p1} in Appendices~\ref{sec:bpmn} and~\ref{sec:rpl-code}, respectively.

%% file: web.tex
	
$\easyrpl$\footnote{\href{https://costa.ls.fi.upm.es/rpltools/easyrpl/clients/web/}{https://costa.ls.fi.upm.es/rpltools/easyrpl/clients/web/}}, is developed using the open-source toolkit
\textsc{EasyInterface}~\cite{fase/DomenechGJS17}, which simplifies the process of
building web-based GUIs. In the following, we are going to describe the web-interface of $\easyrpl$, which provides a comprehensive yet
flexible platform for modelling and analysis.

\begin{figure*}[th!]
\centering
\includegraphics[width=0.9\textwidth]{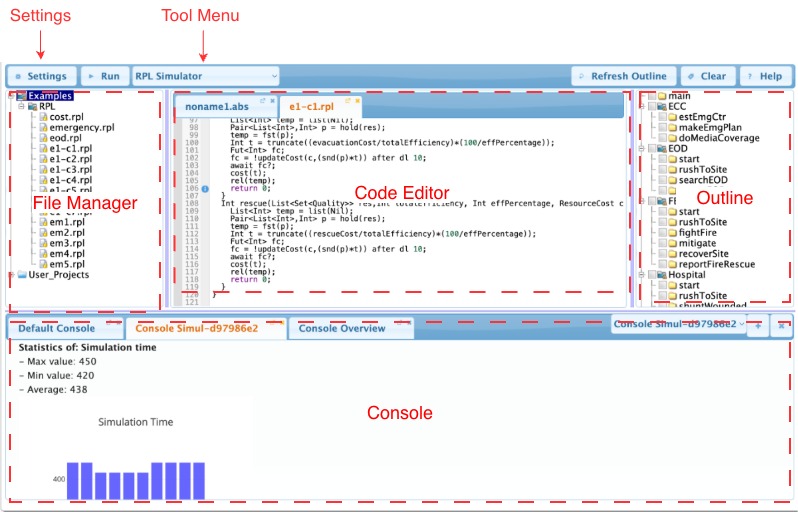}
\caption{Web-interface of $\easyrpl$.}
\label{fig:main}
\end{figure*}


\subsection{Web-based user interface}
Figure \ref{fig:main} illustrates the simple web-based user friendly interface
of $\easyrpl$, displaying its primary components:
\Abs{File Manager}, \Abs{Code Editor}, \Abs{Outline}, \Abs{Console}, \Abs{Settings}, and \Abs{Tool Menu}.
The \Abs{File Manager} provides a structured view of the file organisation,
allowing users to manage their projects and access predefined examples.
The \Abs{Code Editor} serves as the central workspace where users can read and
edit their models.
The \Abs{Outline} section offers an overview of the structure of the
model, displaying essential elements such as class and method
names.
The \Abs{Console} section provides the visualisation of the outputs,
with multiple tabs for organised information.
The \Abs{Settings} section offers users the ability to fine-tune
tool parameters through an intuitive graphical user interface, ensuring precise
control over tool behaviour before execution.
The \Abs{Tool Menu} is a pull down menu allowing the user to select one of the three tools for execution.

%



\subsection{Parameter Configuration}
\label{sec:parameter}

Users can configure different parameters for the \Abs{Simulator} and
\Abs{Peak Resource Analysis} in the~$\easyrpl$:
the parameters resource availability, resource efficiency, and number
of simulations are for the former,
while resources of interest and
quality values of interest are the parameters for the latter.
Additionally, users can specify the number of concurrent instances for both \Abs{Simulator} and \Abs{Peak Resource
  Analysis}. 
%
%
%
%
To configure parameters for these two tools, users need to perform the following two steps:
firstly, using placeholders in the $\rpl$ model to identify the corresponding parameters to be configured; and 
secondly, selecting or customising a profile via the $\easyrpl$
\Abs{Settings} menu.

\begin{figure}[th]
  \centering
  \begin{subfigure}[b]{0.49\textwidth}
    \centering
    \includegraphics[height=3.35cm]{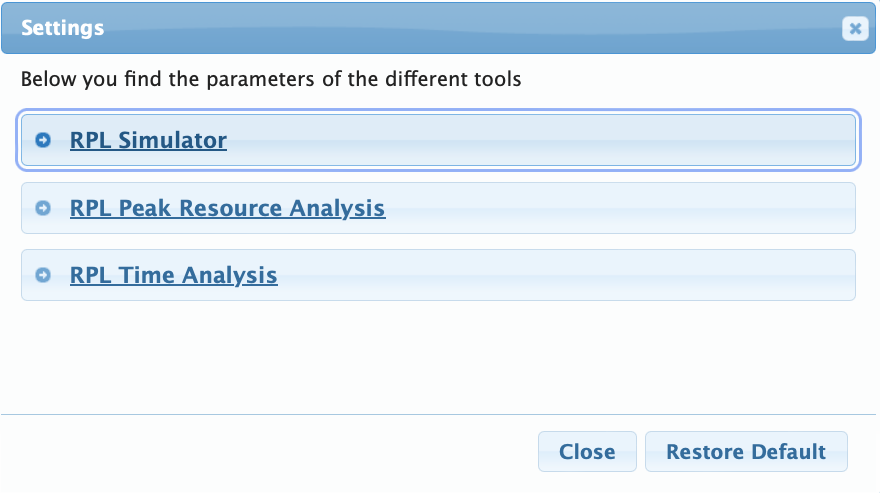}
    \caption{Tool selection.}\label{fig:setting.tool.selection}
  \end{subfigure}
  \begin{subfigure}[b]{0.49\textwidth}
    \centering
    \includegraphics[height=3.35cm]{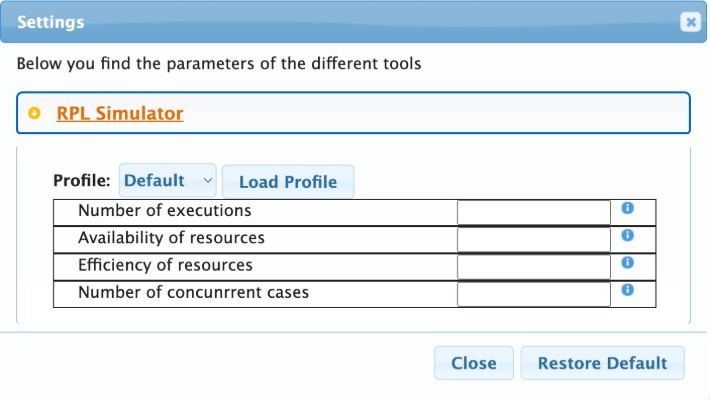}
    \caption{Profile selection.}\label{fig:setting.profile.selection}
  \end{subfigure}%
  \caption{Setting module of $\easyrpl$.}
  \label{fig:setting}
\end{figure}

\paragraph{Placeholders for Configuring Parameters:}
In the $\rpl$ model under consideration, users need to identify
the parameters to be configured in the form of placeholders, which are
%
\$\Abs{AVAILABILITY}, \$\Abs{EFFICIENCY} and \$\Abs{CONC_CASES}
corresponding to the resource availability, efficiency and number of
concurrent cases, respectively.

\paragraph{Profile Selection and Customisation:}
With the parameters specified in the $\rpl$ model as placeholders,
there users then select a profile as an input to the tools.
To select or customise a profile, the users need to %
(1) navigate to the \Abs{Settings} menu, %
(2) select a tool from the list of tools, as shown in
Figure~\ref{fig:setting.tool.selection}, and %
(3) choose a profile from the pre-filled ones, as shown in
Figure~\ref{fig:setting.profile.selection}, that come with preset
values for various configuration parameters.
These profiles are
designed to match common use cases and system capabilities, providing
a quick and efficient way to configure the tool.
Users can also customise these profiles by adjusting the parameters if needed.


\bigskip
Then, upon the execution of the tools, the placeholders in the $\rpl$ model
will be substituted by the values specified for the corresponding
parameters, specifically, 
\$\Abs{AVAILABILITY} is replaced with the value assigned to the availability of
resources. Similarly, \$\Abs{EFFICIENCY} is replaced with the value assigned to
the resource's efficiency percentage, and \$\Abs{CONC_CASES} is replaced with
the value assigned to the number of concurrent cases.


%% file: simulation.tex

This section introduces the simulator in $\easyrpl$. The simulator,
built on top of the ABS compiler, takes an $\rpl$ program, specified
resources, and simulation parameters as input.
These parameters
include the number of concurrent instances, resource availability,
resource efficiency, and the number of simulations.

\begin{figure*}[h!]
	\centering
	\includegraphics[width=0.9\textwidth]{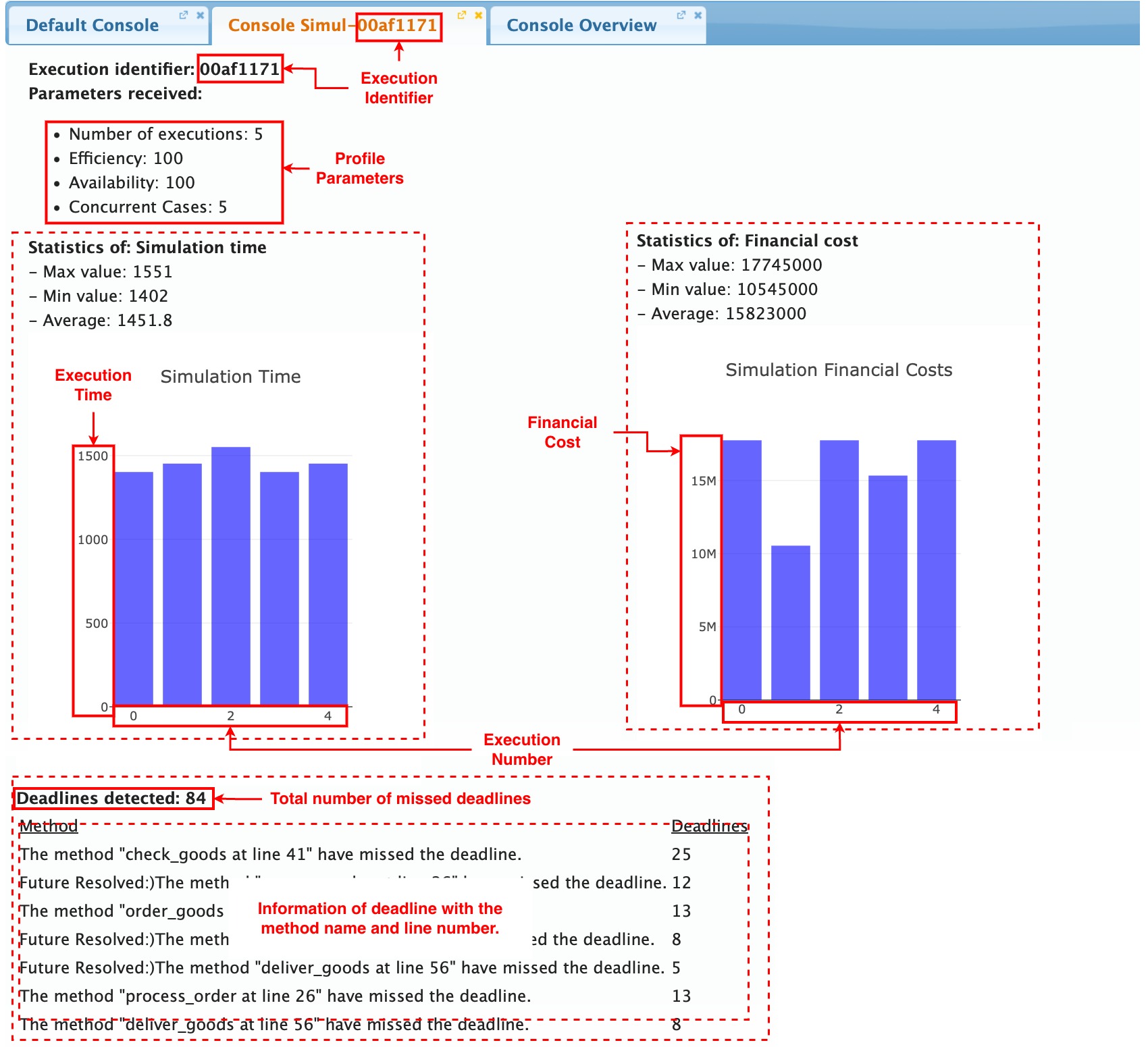}
	\caption{Console of most recent simulation.}
	\label{fig:sim-result}
\end{figure*}

Resource availability and efficiency are key parameters, as in
real-world conditions, resources often do not meet ideal availability
or performance levels due to factors such as maintenance or personnel
inefficiency. Concurrent instances allow users to define how many
instances of a workflow can run simultaneously. This helps in
determining the limits of the current resource pool under different
stress levels and demand scenarios, providing insights into system
performance and behaviour under such conditions. Lastly, the number of
simulations is adjustable, allowing users to determine how many times
the simulation runs to ensure consistent performance across trials.

The simulation process begins with translating the $\rpl$ program into
the corresponding ABS code, which is then compiled using the ABS
compiler.
%
%
To run a simulation, the user needs to
(1) select a particular $\rpl$ model in the \Abs{File Manager}, 
(2) navigate to the \Abs{Tool Menu} and select the $\rpl$ \Abs{Simulator} option,
(3) choose an appropriate profile for the $\rpl$ \Abs{Simulator} from the \Abs{Settings} menu, as illustrated in Figure \ref{fig:setting}. Note that the user can modify the value of the parameters in the profile.
After configuring these settings, the user can (4) initiate the simulation by clicking the \Abs{Run} button. 

%

The results of the simulation are displayed in two separate tabs in
the console area, in addition to the \Abs{Default Console} tab.
The first tab is named \Abs{Console Simul-}\textit{eid}, where
\textit{eid} is an unique identifier of the most recent execution,
while the second tab is named \Abs{Console Overview}. The first tab displays information about the latest execution, while the second tab provides a comparison of all executions that have been performed so far.
\begin{figure*}[h!]
\centering
\includegraphics[width=.9\textwidth]{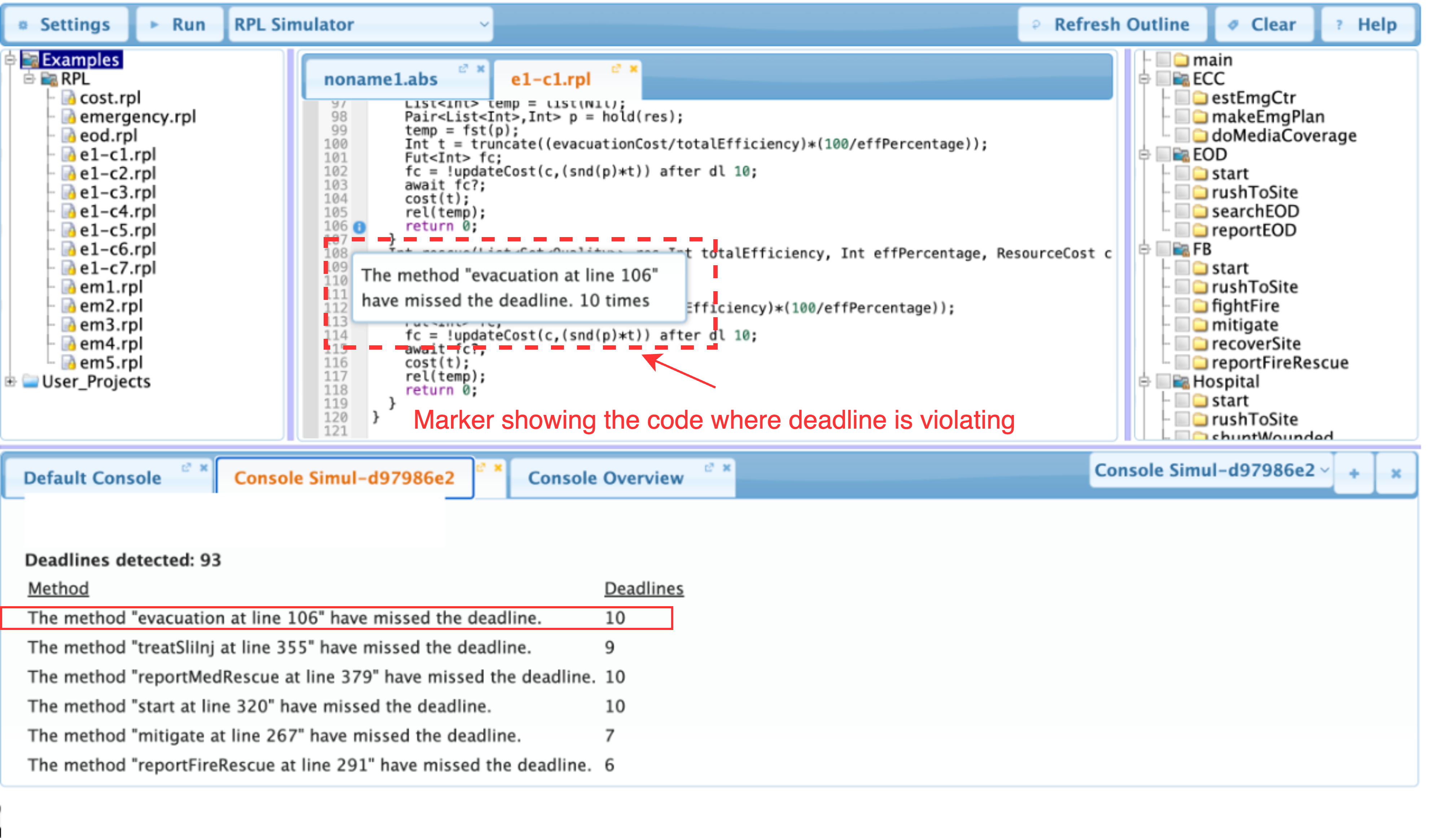}
\caption{Information about deadline violations in the code.}
\label{fig:deadline}
\end{figure*}
Figure~\ref{fig:sim-result} illustrates the representation of the
simulation results right after executing an $\rpl$ model by means of
these two console tabs, where \Abs{00af1171} in the console name
\Abs{Console Simul-00af1171} refers to the identifier of the
execution. 
%
At the top of this console, the identifier value is displayed, followed by the profile parameters used in the simulation. Statistics such as maximum, minimum, and average values for execution time and financial cost are shown along with graphs.
Finally, the total number of deadline violations are presented, along with detailed information about the method name, line number, and number of deadline violations at specific lines. User can also see this information as marked lines in the $\rpl$ code as shown in Figure \ref{fig:deadline}.

\begin{figure*}[h!]
  \centering
  \includegraphics[width=0.9\textwidth]{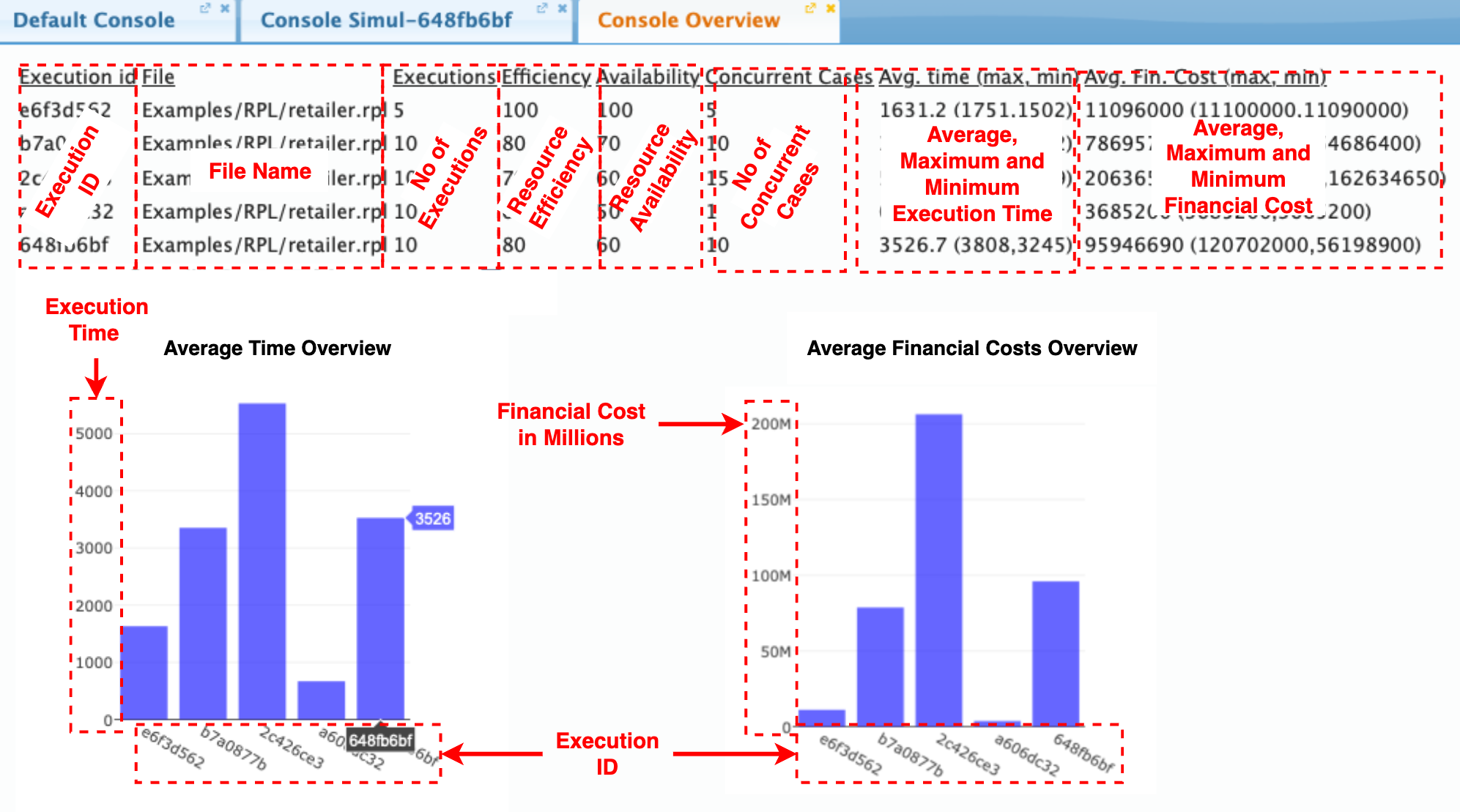}
  \caption{Console Overview.}
  \label{fig:sim-data}
\end{figure*}

Figure~\ref{fig:sim-data} shows the information displayed on \Abs{Console Overview}.
At the top of this console, a table consisting of eight columns is used to present information about various simulations.
Each row in the table shows data from one simulation, and the table grows as new simulations are performed.
The first column represents the unique identifier of the execution.
The second column shows the name of the source file.
The next four columns represent the number of executions, resource efficiency, resource availability, and the number of concurrent instances running simultaneously, respectively.
The last two columns display the average, maximum, and minimum execution times and financial costs for the entire program.

Below the table, we use two bar charts to show the average execution time and average financial costs for all simulations, with the X-axis labelling with the execution identifier and the Y-axis indicating the average execution time and financial cost in millions, respectively.


%% file: peak.tex
As described earlier, cross-organisational workflows typically have a
set of shared resources which are acquired for performing some
specific tasks and, when the tasks are finished, these resources are
released to be used in other tasks.
Given this scenario, one of the most relevant information that an organisation
need to know is the maximum amount of resources of each type that could be
required at the same time. 
To model this behaviour, $\rpl$ includes two statements to
acquire/free resources: \Abs{hold} and \Abs{release}. By means of
\Abs{hold}, the $\rpl$ programmer models the acquisition of a set of
resources needed to perform a specific part of the flow and, with
\Abs{release}, the programmer returns the set of allocated resources previously
acquired.
For example, the implementation of method \Abs{deliver_goods} in Figure~\ref{fig:simple-rpl-2} acquires a set of resources (a \textit{driver} and a
\textit{van}) in Line~\ref{ex:send.hold}
and releases them in Line~\ref{ex:send.release}, respectively.

Traditional cost analysis
techniques~\cite{Albert2012142,albert2011closed,10.1007/978-3-319-28934-2_11,AlbertACGGPR15,10.1007/978-3-662-48288-9_16}
that are defined for obtaining an upper-bound on the amount of cumulative
resources, that is, resources that are never released along the
execution, fall short in approximating such an upper-bound
since resources in the cross-organisational workflows can be allocated concurrently.
%
%
On the contrary, the resource analysis defined in~\cite{AlbertCR18} is able to
obtain a \textit{peak} upper-bound in the form of a cost expression,
which over-approximates the maximum amount of resources that could be allocated
simultaneously along the execution of an ABS~\cite{johnsen2010abs} program.
This peak analysis combines the \textit{may-happen-in-parallel}
analysis~\cite{AlbertFGM15}, which infers the sets of resource allocations that
could occur simultaneously, and the resource analysis~\cite{AlbertACGGPR15} of
ABS programs, which obtains the resources allocated by each set of allocation
sites.

In $\easyrpl$, we apply this peak resource analysis
to a cross-organisational workflow modelled in
$\rpl$ to obtain the peak upper-bound to get the maximum
amount of resources that could be allocated at the same time. To do
so, thanks to the similarities of the concurrency model, we use a
specific transformation to produce an ABS program that preserves the
relevant concurrent properties to the resource analysis:
(1) we keep the classes, object creations and method invocations as they are in
the $\rpl$ program;
(2) we keep the \Abs{await} statements to preserve the concurrent behaviour;
(3) we transform the \Abs{hold} and \Abs{release} $\rpl$ statements into ABS
\Abs{acquire} and \Abs{release} statements to keep the type and cost of the
resource allocated;
(4) we add information to map the $\rpl$ and ABS source code line numbers.

The peak cost analysis uses an \textit{entry} method of the $\rpl$ workflow
model to produce the cost expressions that captures the peak cost of the
program. The entry method is used to define, not only the input parameters of
the cost expression obtained by the analysis, but also the starting point of the
program to be analysed. If the user is interested in analysing the whole model,
this \textit{entry} method could be the \Abs{main} block of the $\rpl$ model,
however, if the user would like to be focused on a specific part of the model,
they could select any other method of the model to perform the analysis. 
Given the entry method, the application of the peak cost analysis on the
transformed ABS program produces a list of upper-bound (UB) expressions that
corresponds to the resources associated to each set of \Abs{hold} statements
which might be allocated
simultaneously starting from the entry method.

\begin{figure*}[t!]
{\small
\[
\begin{array}{c}
\UB_1(p_1, p_2, \ldots, p_n) [l_1, l_2, \ldots, l_m]: \sum \#n_i*(c(\res(\qua_i,\res_i))*k_i+\ldots) \\
\vdots \\
\UB_h(p_1, p_2, \ldots, p_n) [l_1, l_2, \ldots, l_m]: \sum \#n_i*(c(\res(\qua_i,\res_i))*k_i+\ldots) 
\end{array}
\]
}
\caption{Format of upper-bound expressions returned by $\easyrpl$.}
\label{fig:ub-format}
\end{figure*}%
Figure \ref{fig:ub-format} shows the form of the list of UB expressions, each of which sums the resources acquired by a \Abs{hold} statement, returned by $\easyrpl$.
For each expression, we have: $p_1, p_2, \ldots, p_n$,
which are the parameters of the entry method; $l_1, l_2, \ldots, l_m$
corresponds to the line numbers in the $\rpl$ of the \Abs{hold} statement;
$\#n_i$ is the number of times the \Abs{hold} is executed; $\qua_i$ and $\res_i$
correspond, respectively, to the quality and type of resource defined in the
\Abs{hold} statement; and,~$k_i$ is the amount of resources allocated by
the \Abs{hold} statement. 
Additionally, $c(\_)$ and $\res(\_)$ are constructors used to keep in the UBs
the quality and the type of resource acquired in the \Abs{hold} statement.
Observe that a \Abs{hold} statement might acquire a list of pairs
$(\qua_i,\res_i)$
and, all of these pairs will have an addend in the summation
of, at least, one of the expressions.
For example, a \Abs{hold} statement of the
form  
\begin{absexamplesm}
hold(list[set[ResEfficiency(70),Driver], 
         set[ResEfficiency(1500),Van], 
         set[ResEfficiency(1500),Van]])
\end{absexamplesm}
executed in a loop that might iterate $n$ times, produces an addend of the form: 
\[
  n * (c(\res(70,\mathit{Driver})) + c(\res(1500,\mathit{Van})) * 2)
\]

\begin{figure}[!t]
  \centering
  \begin{subfigure}[b]{0.49\textwidth}
    \centering
    \includegraphics[width=.95\linewidth]{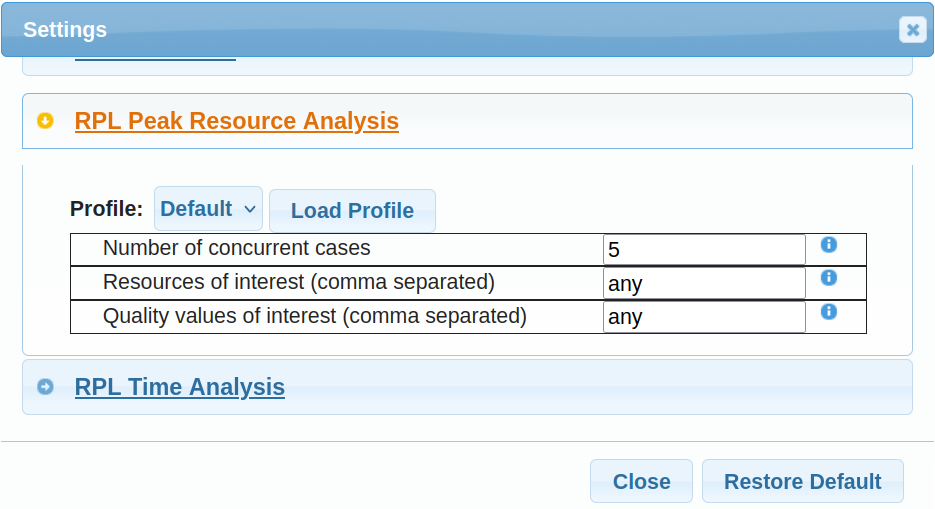}
    \caption{Settings 1.}\label{fig:peak-settings.1}
  \end{subfigure}
  \begin{subfigure}[b]{0.49\textwidth}
    \centering
    \includegraphics[width=.95\linewidth]{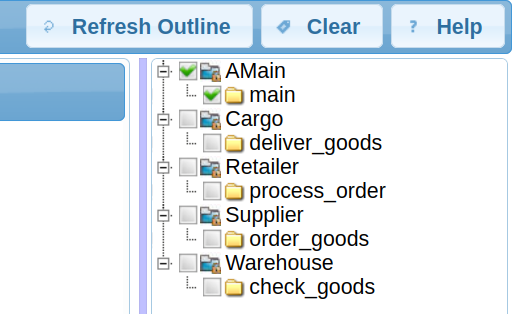}
    \caption{Settings 2.}\label{fig:peak-settings.2}
  \end{subfigure}%
  \caption{\textit{Peak} Resource Analysis settings}
  \label{fig:peak-settings}
\end{figure}

  

Figure~\ref{fig:peak-settings.1} shows the $\easyrpl$ settings to parametrise the
\textit{peak} resource analysis of $\rpl$ programs. This settings configuration
includes: 
\begin{itemize}
\item \textit{Number of concurrent cases} allows the user to specify the value of concurrent instances, indicated by the placeholder \$\Abs{CON\_CASES}, of the workflow model.
\item \textit{Resources of Interest} allows the user to obtain the upper-bound of a set resources of some specific types they are interested in.
\item \textit{Efficiency values of Interest}
  allows the user to obtain the upper-bound for those resources with the set of qualities specified by the user.
\end{itemize}

\medskip

In addition, as mentioned earlier, a method in the model needs to be indicated as an entry point to perform the peak resource analysis, which expresses the cost in terms of the input parameters of the entry method.
To select the entry method, the user first needs to
click on the button \Abs{Refresh Outline} to update the outline with the
methods found in the $\rpl$ code, and then select the method of interest as
entry point, as shown in Figure~\ref{fig:peak-settings.2}. 
The application of the peak analysis to the code shown in
Appendix~\ref{sec:rpl-code} and using the method \Abs{AMain.main} with \emph{five}
concurrent cases returns the expression as follows: 
%
\[
\begin{array}{l}
  \UB(\emph{this}) [\ref{lst:wh.hold.helper}, \ref{lst:supplier.hold.driver.van.helper}, \ref{lst:cargo.hold.driver.van}]:\\
  \begin{array}{rl}
    & 5*(c(\res(70,\mathit{Driver}))+c(\res(1500,\mathit{Van})))+ \\
    & 5*(c(\res(70,\mathit{Driver}))+c(1500,\mathit{Van})))+ \\
    & 5*(c(\res(70,\mathit{Driver}))+c(\res(1500,\mathit{Van}))+ \\
    & ~~~~~~\!c(\res(50,\mathit{Helper})))\ + \\
    & 5*c(\res(50,\mathit{Helper}))
  \end{array}
\end{array}    
\]
%
%
As we can see from the expression that the analysis returns only one
UB with
Lines~\ref{lst:wh.hold.helper},~\ref{lst:supplier.hold.driver.van.helper}
and~\ref{lst:cargo.hold.driver.van} (all lines with \Abs{hold}
statements), which implies that all resources allocated by the model
could be in use at the same time.

If the user is interested in a particular resource, they can specify
it in the parameter \emph{Resources of Interest} in the \Abs{Settings}
menu in Figure~\ref{fig:peak-settings.1}.  For instance, setting this
parameter to \Abs{Van}, the analysis returns the expression:
$$\UB(\mathit{this}) [\ref{lst:wh.hold.helper}, \ref{lst:supplier.hold.driver.van.helper}, \ref{lst:cargo.hold.driver.van}]: 15$$
which says that, given five concurrent cases, the company will need
$15$ \Abs{Van}s to perform all tasks concurrently.
Similarly, the user can specify the efficiencies of interest by
setting the parameter {\codesize\sffamily Quality values of interest} in the
\Abs{Settings} menu in Figure~\ref{fig:peak-settings.1}.  For
instance, setting the efficiency to $50,120$ we get the following UB
$$5*2+5*2+5*1\ ,$$%
which shows that the maximum amount of resources of these efficiencies
demanded at the same time is $25$.

\begin{figure}[t]
  \centering
  \includegraphics[width=\linewidth]{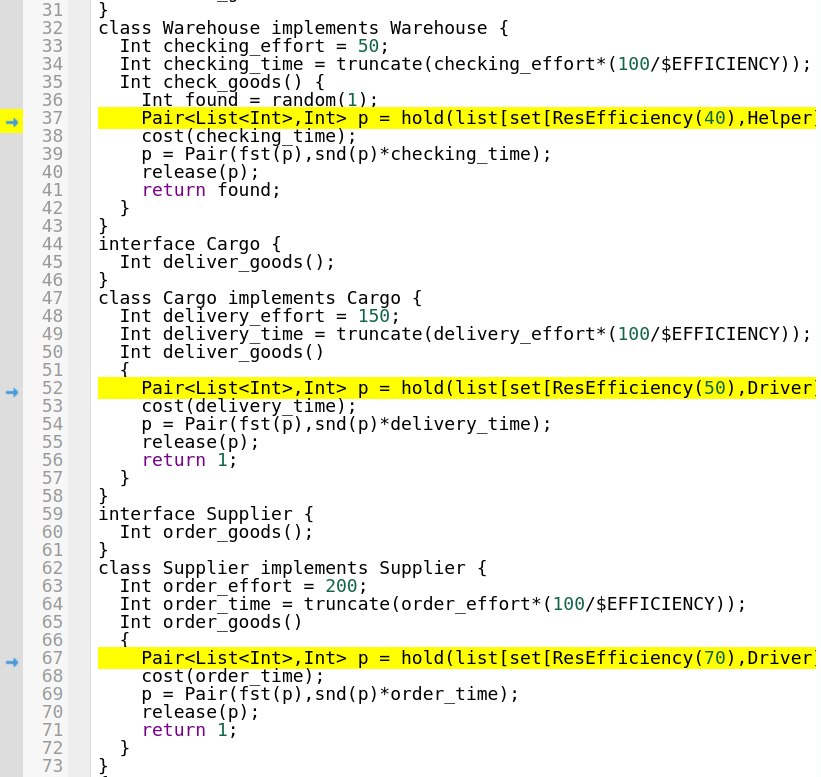}  
  \vspace{10pt}
  \caption{\textit{Peak} Resource Analysis -- Code Highlights }
\label{fig:peak.highlights}
\end{figure}

The user interface also includes the possibility of interacting with the source
code: when the peak analysis is completed, some arrows appear to the
left of the source code lines with \Abs{hold} statements.
By clicking on one of the arrows,
the interface highlights in the $\rpl$ model all the lines with a \Abs{hold} statement that could
allocate resources simultaneously as the selected line, as shown in Figure~\ref{fig:peak.highlights}.
%

%

Interestingly, to illustrate how our peak analysis can obtain different
set of resources, if we select as entry method \Abs{process\_order}, that is, if
the model does not have any concurrency in processing the orders, we get the
expression as follows: 
\[
\begin{array}{l}
  \UB(\mathit{this,wr,cr,sup}) [L\ref{lst:wh.hold.helper}]: c(\res(50,\mathit{Helper})) \\[.3em] 
  \UB(\mathit{this,wr,cr,sup}) [L\ref{lst:supplier.hold.driver.van.helper}]:\\
  ~~c(\res(70,\mathit{Driver}))+ c(\res(1500,\mathit{Van}))+\\
  ~~c(\res(50,\mathit{Helper})) \\[.3em] 
  \UB(\mathit{this,wr,cr,sup}) [L\ref{lst:cargo.hold.driver.van}]: \\
  ~~c(\res(70,\mathit{Driver}))+c(\res(1500,\mathit{Van})) \\[.3em] 
  \UB(\mathit{this,wr,cr,sup}) [L\ref{lst:cargo.hold.driver.van}]:\\
  ~~c(\res(70,\mathit{Driver}))+c(\res(1500,\mathit{Van}))
  
\end{array}
\]
%
As we can see in the code, all resources are allocated and released in the
same method, thus, this list of UB expressions is capturing this notion as two
\Abs{hold} statements could have allocated resources at the same time. Note that
the resources allocated in Line~\ref{lst:supplier.hold.driver.van.helper} corresponds to a \Abs{hold} statement with three
different resources and, interestingly, the UBs in Line~\ref{lst:cargo.hold.driver.van}  indicates that this
line could be executed twice, but the resources allocated by the two executions will 
not be simultaneously in~use.

%% file: time.tex
In addition to the definition and allocation of the available resources, $\rpl$
also allows the users to specify the amount of time needed to perform 
specific parts of a cross-organisational workflow using the statement
\Abs{cost(t)}, and with which time advances for~$t$ time units in the $\rpl$ workflow model. 

The work presented in~\cite{ALI2023102896} extends the time analysis proposed in~\cite{LaneveLPR19} to analyse workflows modelled in $\rpl$.
The idea is based on the use of synchronisation sets containing objects whose executions might be related to create a set of cost equations, which captures the execution time of the model.  The solution to this set of cost equations 
represents an upper-bound on the time taken for
executing the model. 
The main difference between the analyses presented in~\cite{LaneveLPR19}
and~\cite{ALI2023102896} is in the input language and the features included:
the one in~\cite{LaneveLPR19} is a basic language intended to serve
as a compiler target and includes neither conditional statements nor while loops.
On the contrary, $\rpl$ in the latter is a full modelling language which is
intended to allow the users to define cross-organisational workflows. 
Another main difference between the time analysis of~\cite{LaneveLPR19}
and the analysis of~\cite{ALI2023102896} is in the generation and the resolution
of the cost equations.
While the latter expresses the WCET in the form of a set of cost equations without coupling to any constraint solvers,
the former has been coupled with 
PUBS~\cite{AlbertAGP11a}, an off-the-shelf equation solver, which includes a size analysis and a
maximisation process, solving the cost equations and obtaining a
closed-form upper-bound expressed in terms of the input parameters.

To calculate a closed-form upper-bound of an $\rpl$ model with $\easyrpl$, we first generate the corresponding ABS programs on which we apply the time analysis defined in~\cite{LaneveLPR19}.
%
%
To handle conditional statements in $\rpl$ models, $\easyrpl$ produces
as many ABS programs as the number of potential execution paths the $\rpl$ program could
have.
As a result, the WCET of the $\rpl$ model is the maximum of the WCET of the produced ABS programs.
The transformation for the time analysis is similar to the one in Section~\ref{sec:peak}:
it preserves the concurrent features of the program by keeping object creations, method invocations and synchronisations,
but it transforms the $\rpl$ \Abs{cost} statements into \textbf{\Abs{wait}} statements in the ABS program, on which we apply
the PUBS solver, returning a
closed-form upper-bound of the execution time of the $\rpl$ model expressed in terms
of the input parameters of the program or in terms of the maximum value taken by
the actor fields in case the upper-bound depends on them. 

Note that, the time analysis in both~\cite{LaneveLPR19} and~\cite{ALI2023102896} cannot handle loops but only recursions, 
and the loops used in $\rpl$ models is for iterating simulation and does not involve any \Abs{cost} statement; thus, they are removed during the translation from $\rpl$ to ABS for time analysis, whose correctness is not affected by such removal.

The application of the \textit{time analysis} to the running example returns
that the worst-case execution time is the maximum between the following
two expressions: 

\[
  \begin{array}{rl}
    
    \UB_1 \main(\mathit{this}) = & \mathit{Warehouse.checking\_time}\\
    &+\mathit{Cargo.delivery\_time }\\[.3em] 
    \UB_2 \main(\mathit{this}) = & \mathit{Warehouse.checking\_time}\\
    &+ \mathit{Supplier.order\_time}\\
    & +\mathit{ Cargo.delivery\_time}
\end{array}
\]

As shown in the model shown in Figure~\ref{fig:simple-rpl-1}, the code
of \Abs{process_order} includes an if-then-else statement that produces two
possible \alt programs that return two different upper-bound expressions.
Note that the expression included into the \Abs{cost} expressions
depends on values stored in different fields and the time analysis
captures it in the cost expression. Note also that, as we have already
mentioned, the resulting is not considering the loop in the \Abs{main}
method.


%% file: related.tex
$\easyrpl$ has three distinct tools: simulation, resource peak analysis, and time analysis. In this section, we present the work related to each of these domains to provide an understanding of their relevance.

\medskip
\noindent
\textbf{Simulation.}
Simulation-based tools and techniques have gained attention for their ability to assess resource policies, evaluate performance, and address real-world system issues~\cite{ahuja1985simulation,LI2009286}.
In~\cite{10.1007/978-3-030-26643-1_6}, a method employing Structural Causal Models (SCM)~\cite{pearl2009causality} has been proposed to reason about cause-effect assumptions. This approach utilises the online BIMP tool\footnote{\href{https://bimp.cs.ut.ee/simulator}{https://bimp.cs.ut.ee/simulator}} to simulate event logs based on selected Business Process Model and Notation (BPMN)~\cite{10.1007/978-3-031-12441-9_1} specifications.
Additionally, in~\cite{10.1007/978-3-031-07472-1_11}, an approach integrated into Visual Miner~\cite{10.1007/978-3-319-15895-2_26} within the ProM framework~\cite{10.1007/11494744_25} analyses causality in business processes. Despite their contributions, these studies may not offer formal support for modelling cross-organisational workflows or provide the depth of insights necessary for decision-making in collaborative environments.
In contrast, the work presented in~\cite{DEBOIS20161295} applies Dynamic Condition Response (DCR), a constraint-based modelling notation, to develop emergency response plans for railways. While DCR offers flexibility in modelling various emergency scenarios through essential rules or conditions, it currently does not support the concurrent execution of multiple independent events.
UPPAAL~\cite{behrmann2006uppaal} is being employed for modelling, simulating, and verifying workflows represented in timed automata. SWEET \cite{Ermedahl163045} has the capability to generate models using UPPAAL syntax \cite{Sundmark121024}, whose constructs
are designed to be very small and simple~\cite{gustavsson_et_al:OASIcs:2010:2830}.
In contrast, $\easyrpl$ allows simulate workflows modelled in $\rpl$, that facilitates the integration of cross-organisational workflows involving shared resources and task dependencies. Unlike other methodologies, this approach enables decision-makers not only to foresee consequences within individual workflows but also to gain insights into the broader implications across all collaborating workflows.

\medskip
\noindent
\textbf{Resource Peak Analysis.}
Most research on cost analysis of sequential imperative programs has primarily focused on resource usage in terms of memory consumption with garbage collection~\cite{ALBERT20131427,10.1007/978-3-642-37036-6_32}.
Furthermore, resource analysis through the use of \textbf{\Abs{malloc}} and \textbf{\Abs{free}} instructions (analogous to our \Abs{hold} and \Abs{release} statements), has been extensively studied for sequential programs \cite{10.1145/2737924.2737955,5351120}.
In \cite{hoffmann2015automatic}, such analyses have been extended to parallel settings but do not deal with concurrency.
The static analysis method introduced in~\cite{GARCIA201727} aimed at determining the peak cost of live virtual machines in a concurrent programming language that uses explicit \Abs{acquire} and \Abs{release} operations for virtual machines.
Their technique primarily focuses on calculating the count of virtual machines, assigning one virtual machine per acquire instruction. In contrast, the approach we used in $\easyrpl$ extends beyond this by handling multiple types of resources concurrently.

\medskip
\noindent
\textbf{Time Analysis.}
Various tools have been developed to over-approximate the execution time of systems. For example, Chronos~\cite{chronos} is a static worst-case execution time (WCET) analysis tool specialised for embedded systems, modelling features like branch prediction, instruction cache, and out-of-order execution.
Another tool, OTAWA~\cite{otawa}, provides a framework of C++ classes for static analysis of machine code and WCET calculation. It supports the development of custom analyses for new hardware or experimental needs.
However, these existing tools are criticised for lacking advanced control flow mechanisms necessary for workflow modelling. In contrast, $\easyrpl$ supports a more expressive language enabling modelling of workflows with task dependencies, synchronisation, and shared resources.


%% file: example-code.tex
\begin{figure*}[h!]
\begin{minipage}[h]{1\linewidth}
\begin{absexamplesm}[name=runningex]
// Model of supply chain in $\rpl$
module Retail;
interface Retailer {
$~~$Int process_order(Warehouse wr,Cargo cr,Supplier sup);
}
class Retailer implements Retailer {
$~~$Int found = 0; 
$~~$Int process_order(Warehouse wr,Cargo cr,Supplier sup)
$~~${
$~~$$~~$Int sent = 0;
$~~$$~~$Fut<Int> f1; Fut<Int> f2; Fut<Int> f3;
$~~$$~~$f1 = !check_goods(wr) after dl 10;
$~~$$~~$await f1?;
$~~$$~~$found = f1.get;
$~~$$~~$if(found == 1)
$~~$$~~${
$~~$$~~$$~~$f2 = !deliver_goods(cr) after dl 170;
$~~$$~~$$~~$await f2?;
$~~$$~~$$~~$sent = f2.get;
$~~$$~~$} else {
$~~$$~~$$~~$f2 = !order_goods(sup) after dl 220;
$~~$$~~$$~~$f3 = !deliver_goods(cr) after f2 dl 170;
$~~$$~~$$~~$await f3?;
$~~$$~~$$~~$sent = f3.get;
$~~$$~~$}
$~~$$~~$return sent;
$~~$}
}
\end{absexamplesm}
\end{minipage}

\begin{minipage}[h]{1\linewidth}
\begin{absexamplesm}[name=runningex]
interface Warehouse {
$~~$Int check_goods();
}
class Warehouse implements Warehouse {
$~~$Int checking_effort = 50;
$~~$Int checking_time = truncate(checking_effort*(100/$\$$EFFICIENCY));
$~~$Int check_goods() {
$~~$$~~$Int found = random(1);
$\label{lst:wh.hold.helper}~~$$~~$Pair<List<Int>,Int> p = hold(list[set[ResEfficiency(50),Helper]]);
$~~$$~~$cost(checking_time);
$~~$$~~$p = Pair(fst(p),snd(p)*checking_time);
$~~$$~~$release(p);
$~~$$~~$return found;
$~~$}
}
\end{absexamplesm}
\end{minipage}

\begin{minipage}[h]{1\linewidth}
\begin{absexamplesm}[name=runningex]
interface Cargo {
$~~$Int deliver_goods();
}
class Cargo implements Cargo {
$~~$Int delivery_effort = 150;
$~~$Int delivery_time = truncate(delivery_effort*(100/$\$$EFFICIENCY));
$~~$Int deliver_goods()
$~~${
$\label{lst:cargo.hold.driver.van}~~$$~~$Pair<List<Int>,Int> p = hold(list[set[ResEfficiency(70),Driver],set[ResEfficiency(1500),Van]]);
$~~$$~~$cost(delivery_time);
$~~$$~~$p = Pair(fst(p),snd(p)*delivery_time);
$~~$$~~$release(p);
$~~$$~~$return 1;
$~~$}
}
\end{absexamplesm}
\end{minipage}
\caption{Workflow model of running example in $\rpl$ -- part I}
\label{fig:p1}
\end{figure*}

\begin{figure*}[ht]\ContinuedFloat
\begin{minipage}[h]{1\linewidth}
\begin{absexamplesm}[name=runningex]
interface Supplier {
$~~$Int order_goods();
}
class Supplier implements Supplier {
$~~$Int order_effort = 200;
$~~$Int order_time = truncate(order_effort*(100/$\$$EFFICIENCY));
$~~$Int order_goods() {
$~~$$~~$Pair<List<Int>,Int> p =
$\label{lst:supplier.hold.driver.van.helper}~~$$~~$hold(list[set[ResEfficiency(70),Driver],set[ResEfficiency(1500),Van],set[ResEfficiency(50),Helper]]);
$~~$$~~$cost(order_time);
$~~$$~~$p = Pair(fst(p),snd(p)*order_time);
$~~$$~~$release(p);
$~~$$~~$return 1;
$~~$}
}
\end{absexamplesm}
\end{minipage}

\begin{minipage}[h]{1\linewidth}
\begin{absexamplesm}[name=runningex]
{
$~~$Int counter = 1;
$~~$Int max = $\$$CONC_CASES;
$~~$List<Fut<Int>> fl = Nil;
$~~$Retailer ret = new Retailer();
$~~$Warehouse wr = new Warehouse();
$~~$Cargo cr = new Cargo();
$~~$Supplier sup = new Supplier();
$~~$while (counter <= max)
$~~${
$~~$$~~$Fut<Int> f;
$~~$$~~$f = !process_order(ret,wr,cr,sup) after dl 400;
$~~$$~~$fl = appendright(fl,f);
$~~$$~~$counter = counter + 1;
$~~$}
$~~$while(!isEmpty(fl))
$~~${
$~~$$~~$Fut<Int> f = head(fl);
$~~$$~~$await f?;
$~~$$~~$fl = tail(fl);
$~~$}
}
\end{absexamplesm}
\end{minipage}

\begin{minipage}[h]{1\linewidth}
\begin{absexamplesm}[name=runningex]
Resources:
Van,1500,5000,1
Van,1500,5000,1
Van,1500,5000,1
Van,1500,5000,1
Van,1500,5000,1
Van,1500,5000,1
Van,1500,5000,1
Van,1500,5000,1
$\$$
Driver,70,1000,1
Driver,70,1000,1
Driver,70,1000,1
Driver,70,1000,1
Driver,70,1000,1
Driver,70,1000,1
Driver,70,1000,1
Driver,70,1000,1
$\$$
Helper,50,500,1
Helper,50,450,1
Helper,50,450,1
Helper,50,500,1
\end{absexamplesm}
\end{minipage}
\caption{Workflow model of running example in $\rpl$ -- part II}
\label{fig:p2}
\end{figure*}
